\documentclass[a4paper,fleqn,usenatbib,letters]{mnras}

\usepackage{newtxtext,newtxmath}
\usepackage{txfonts}

\usepackage[T1]{fontenc}
\usepackage{ae,aecompl}

\usepackage{graphicx}	
\usepackage{epstopdf}
\usepackage{amsmath}	
\usepackage{amssymb}	
\usepackage{mathtools}	

\title[Orbital motion of CW Leo]{Evidence for orbital motion of CW Leonis from ground-based astrometry}

\author[A. Sozzetti et al.]{
A. Sozzetti,$^{1}$\thanks{E-mail: sozzetti@oato.inaf.it}
R.L. Smart,$^{1,2}$
R. Drimmel,$^{1}$
P. Giacobbe$^{1}$
and M.G. Lattanzi$^{1}$
\\
$^{1}$INAF - Osservatorio Astrofisico di Torino, Via Osservatorio 20, I-10025 Pino Torinese, Italy\\
$^{2}$Centre for Astrophysics Research, University of Hertfordshire, College Lane, Hatfield, AL10 9AB\\
}

\date{Accepted XXX. Received YYY; in original form ZZZ}

\pubyear{2017}

\begin{document}
\label{firstpage}
\pagerange{\pageref{firstpage}--\pageref{lastpage}}
\maketitle

\begin{abstract}
Recent ALMA observations indicate that CW Leo, the closest carbon-rich AGB star to the Sun, might have a low-mass stellar companion. 
We present archival ground-based astrometric measurements of CW Leo obtained within the context of the Torino Parallax Program and with $>6$ yr 
($1995 - 2001$) of time baseline. The residuals to a single-star solution show significant curvature, and they are strongly correlated with the well-known 
$I$-band photometric variations due to stellar pulsations. We describe successfully the astrometry of CW Leo with a variability-induced motion (VIM) + acceleration model. 
We obtain proper motion and parallax of the center-of-mass of the binary, the former in fair agreement with recent estimates, the latter at the 
near end of the range of inferred distances based on indirect methods. The VIM + acceleration model results allow us to derive a companion mass in agreement 
with that inferred by ALMA, they point towards a somewhat longer period than implied by ALMA, but are not 
compatible with much longer period estimates. These data will constitute a fundamental contribution towards the full understanding of the orbital architecture 
of the system when combined with Gaia astrometry, providing a $\sim 25$\,yr time baseline. 

\end{abstract}

\begin{keywords}
stars: AGB and post-AGB -- astrometry -- circumstellar matter -- binaries: general -- stars: individual: CW Leo -- stars: late-type
\end{keywords}



\section{Introduction}

The brightest extrasolar object in the sky at 5 $\mu$m, CW Leo (also known as the infrared source IRC+10216) 
is the closest carbon-rich, long-period pulsating variable, asymptotic giant branch (AGB) star to the Sun, with an estimated distance between 
120 and 150\,pc (e.g., \citealt{Menshchikov2001}; \citealt{Groenewegen1998}, \citeyear{Groenewegen2012}, and references therein). Embedded in an expanding cocoon of material 
originating from the star itself due to the mass-loss process characteristic of the AGB evolutionary phase, this remarkable object has been 
the subject of extensive studies across a wide range of wavelengths, aimed at understanding its circumstellar environment via detection of new molecules and 
the characterization of the kinematic, dynamical, chemical, structural, and dust properties of the material (e.g., see references in 
\citealt{Menten2012,Kim2015}; \citealt{Decin2015} (D15 hereafter); \citealt{Stewart2016}). 

One peculiar aspect of the environment surrounding CW Leo is that, while the overall shape of the circumstellar mass-loss envelope (CSE) skulpted by 
CW Leo's winds appears fairly spherical, high spatial resolution observations have unveiled a variety of structures at different spatial scales and in different
wavelength regions: These include spiral shells and dust clumps at subarcsecond scale, indications of bipolar structure at arcsecond scales, and 
the presence of multiple non-concentric shells in the outer wind. The most recent observational evidence of the complex morphology and kinematics of the inner 
regions of CW Leo's CSE has provided circumstantial evidence for the presence of a binary companion. Indirect inferences on the companion properties (mass, 
orbital separation, eccentricity) are made through comparison of the observations with models of the effects a binary companion has on the wind envelope structure. 
Such inferences indicate the companion likely being a low-mass K or early M dwarf, while the orbital period is relatively uncertain (from tens to several hundreds of years), 
and its orbital geometry is even less constrained. 

Direct observational evidence for binarity in AGB stars such as CW Leo cannot be readily obtained. In particular, the very high luminosity ($\sim10^4$ L$_\odot$) 
of CW Leo and its complex, dynamically evolving dusty envelope make it very difficult to directly detect a low-mass main-sequence companion 
\citep{Kim2015,Stewart2016}. 
It is a challenge to utilize photometric variability monitoring and radial-velocity measurements, 
because the strong variability intrinsic to CW Leo's pulsating atmosphere can potentially mask the corresponding variability due to a companion. 
In this Letter we present ground-based relative astrometry of CW Leo gathered within the context of the Torino Observatory Parallax Program (\citealt{Smart2003}, hereafter TOPP) 
covering a time baseline of over 6 years. Positional measurements of CW Leo allow to derive a direct distance determination at the near end of the range of 
published indirect estimates. The astrometric residuals exhibit significant curvature, which we interpret as evidence of orbital motion due to a companion. 
The TOPP astrometric measurements also allow for detection of variability-induced motion (VIM), that we use to put additional constraints on the orbital architecture 
of the binary system.

\section{Observations and data reduction}

\begin{figure}
\centering
   \includegraphics[width=0.95\columnwidth]{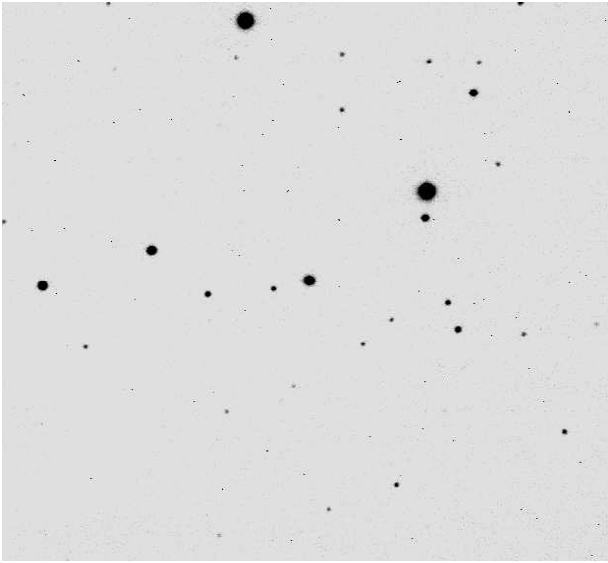}
   \caption{The base CCD image from 1997.26 centered on CW Leo. The field of view is 10 $\times$ 9\,arcmin, North is orientated up and East is left.}
   \label{fig1}
\end{figure}

CW Leo was observed as part of TOPP from January 1995 to April 2001. Observations were all carried out on the 1.05m reflecting telescope which is a
scaled-down version {of the 1.55-m Kaj Strand Astrometric Reflector at the USNO Flagstaff Station \citep{Strand1966}.} The detector used was an EEV
CCD05-30 1296x1152 @ 15 $\mu$m pix$^{-1}$ constructed by the Astromed company which provided a pixel scale of $0.47^{\prime\prime}$ and a field of view of 
$10^\prime\times9^\prime$. All parallax observations were carried out in the Cousins $I$ filter. An image of the CW Leo field is presented in Fig. \ref{fig1}.

The full procedures, treatment and reduction for the TOPP program is described in Smart et al. (\citeyear{Smart1999,Smart2003}). Here we just briefly list
the main steps.  All images are flat fielded using nightly sky flats and bias-corrected using image overscan regions. All objects in the field are
found and centroided using inhouse gaussian profile fitting software. We then choose a good frame in the middle of the sequence and using anonymous stars in
common with other frames we astrometrically adjust the other frames to this frame. Selection of the reference stars and usable frames is then carried out
by eliminating stars with high across season errors, or eliminating frames that have a small number of common stars, or positional residuals larger 
than three times the average frame residual. Both TOPP astrometry and photometry of CW Leo utilized in the analysis are provided in machine-readable
form in the online Journal. 

A given sequence is iterated to obtain proper motions and parallaxes for all objects and the above criteria are applied until the estimated parallax of the
target changes by less than 1\%. We then apply a correction from the calculated relative parallax to an astrophysically useful absolute parallax
using estimates of the field star distances from the \citet{Mendez1998} galaxy model. The error of the final 
parallax is found by adding in quadrature to the formal error $33\%$ of the relative to absolute correction. 

By default this program assumes the target is a single star, we do not attempt to model at this stage any possible binary nature. In table \ref{sol}
we list the parameters of the CW Leo sequence using this single-star astrometric solution. The solution has a reduced 
chi-square $\chi^2_r\sim42$. The excess residuals indicate the presence of unmodeled effects that are not captured by the single-star model.

\begin{table}
\centering
\caption{Log of observations and single-star solution for CW Leo}
\begin{tabular}{lr}
\hline  \\
 $\alpha, \delta$ ~~ (J2000)          & 9:47:57.3, +13:16:43.5 \\ 
 Mean Epoch [yr]                             &       1997.2644 \\ 
 $\pi_\mathrm{abs}$ [mas]                &  10.79 $\pm$  4.60 \\ 
 $\mu_{\alpha^\star}$  [mas yr$^{-1}$]             &  37.50 $\pm$  1.54 \\ 
 $\mu_{\delta}$  [mas yr$^{-1}$]            &  30.22 $\pm$  2.02 \\ 
 Rel. to abs. correction [mas]  & 1.72 \\ 
 Obs. start, timespan [yr]    & 02/01/1995, 6.30 \\ 
 N. of reference stars, obs. & 31, 139 \\ 
\hline 
\label{sol}
\end{tabular}
\end{table} 

\begin{figure}                                          
   \includegraphics[width=0.95\columnwidth]{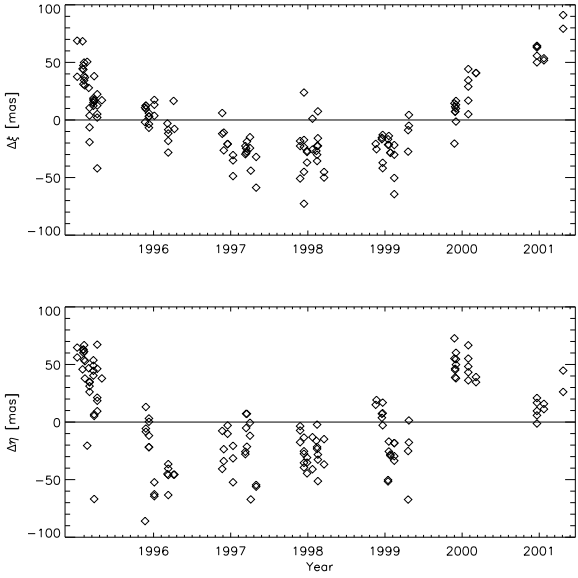}
   \caption{Residuals in standard coordinates $\xi,\eta$ to a simple single-star solution for all 
   observations not rejected by the standard pipeline rejection criteria. }
   \label{fig2}
\end{figure}

In Fig. \ref{fig2} we plot the residuals of the CW Leo fit over the observational sequence. The residuals exhibit a distinct curvature, hinting at a 
period exceeding the $\sim6$\,yr of the observational campaign. 
Prompted by the recent analysis of D15, we further analyzed the data to see if a binary solution could explain the observed residuals. 

\section{Results}

The evidence of a long-term trend in the residuals to the standard five-parameter solution is statistically solid, 
based on the GLS periodogram analysis \citep{Zechmeister2009}, with a bootstrap-based false-alarm probability FAP $\sim10^{-40}$. 
We then attempted to describe the data in terms of a seven-parameter model that includes derivatives of the 
proper motion. The best-fit solution is presented in Table \ref{solacc}. The reduced chi-square of the acceleration solution ($\chi^2_r\sim19$) 
is improved with respect to the five-parameter model value. We quantify the likelihood that acceleration in the astrometric data of CW Leo 
is detected based on an $F$-test (in essence, a likelihood-ratio test) that evaluates the significance of the decrease of the $\chi^2$ resulting from 
the addition of two parameters. The $F$-test gave a probability of $\sim10^{-26}$ that the single-star solution is a better description of the data, 
thus the accelerated motion of CW Leo is considered detected with high statistical confidence. 

\begin{table}
\centering
\caption{The acceleration solution for CW Leo. Uncertainties are derived via bootstrap method.}
\begin{tabular}{lr}
\hline  \\
 offset in $\alpha$ [mas]         & $21.05\pm1.31$  \\ 
 offset in $\delta$ [mas]                            &  $30.69\pm1.34$ \\ 
 $\mu_{\alpha^\star}$  [mas yr$^{-1}$]             &  30.41 $\pm$  1.47 \\ 
 $\mu_{\delta}$  [mas yr$^{-1}$]            &  23.73 $\pm$  1.45  \\ 
 $\pi_\mathrm{abs}$ [mas]                &  12.33 $\pm$  1.90\\   
 $\dot \mu_{\alpha^\star}$  [mas yr$^{-2}$]             &  8.34 $\pm$  1.24\\ 
 $\dot \mu_{\delta}$  [mas yr$^{-2}$]            &  8.57 $\pm$  1.23 \\ 

\hline 
\end{tabular}
\label{solacc}
\end{table} 

\begin{figure}
 \begin{center}
$\begin{array}{c}
    \includegraphics[width=0.95\columnwidth]{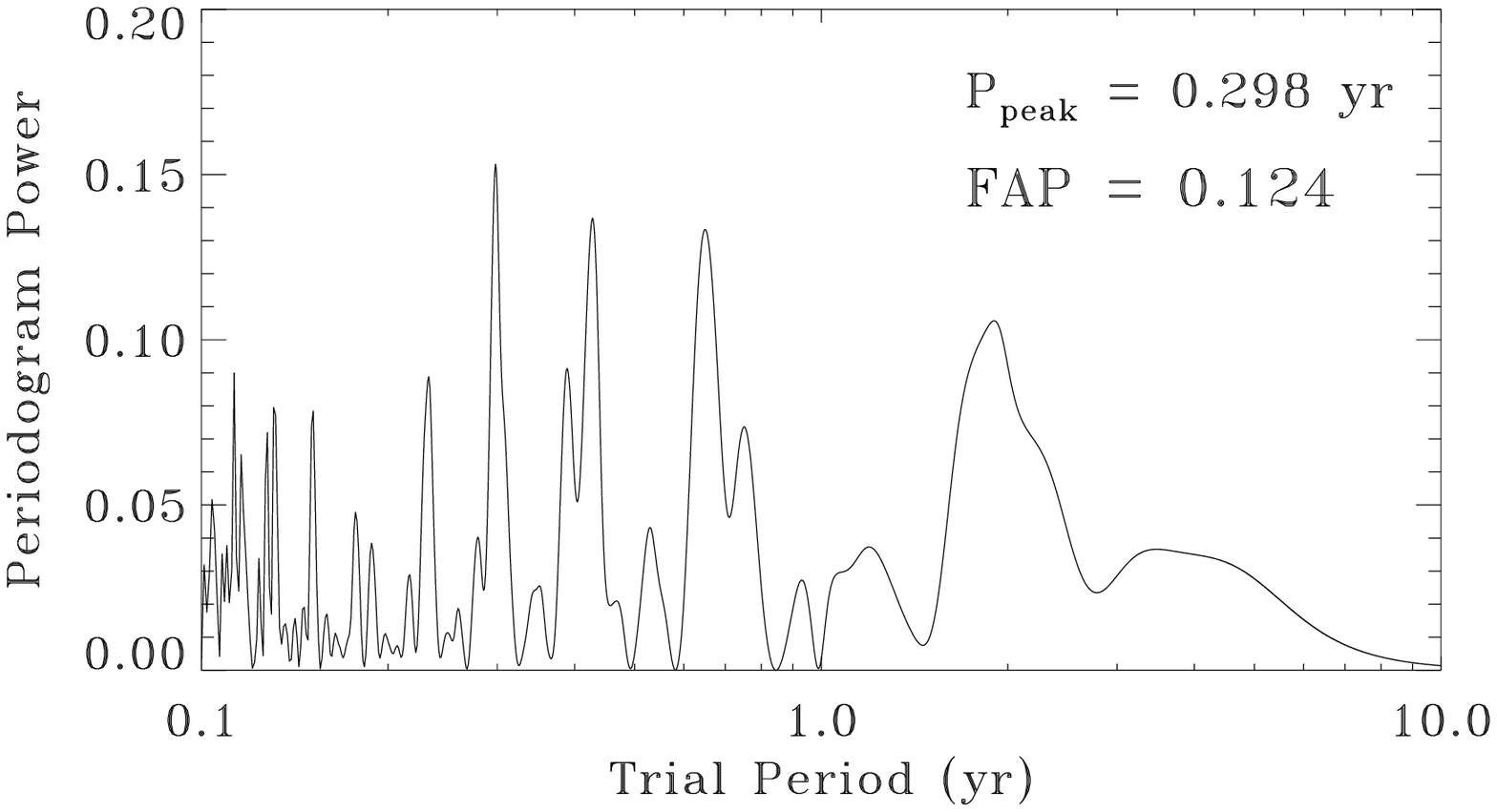} \\
    \includegraphics[width=0.95\columnwidth]{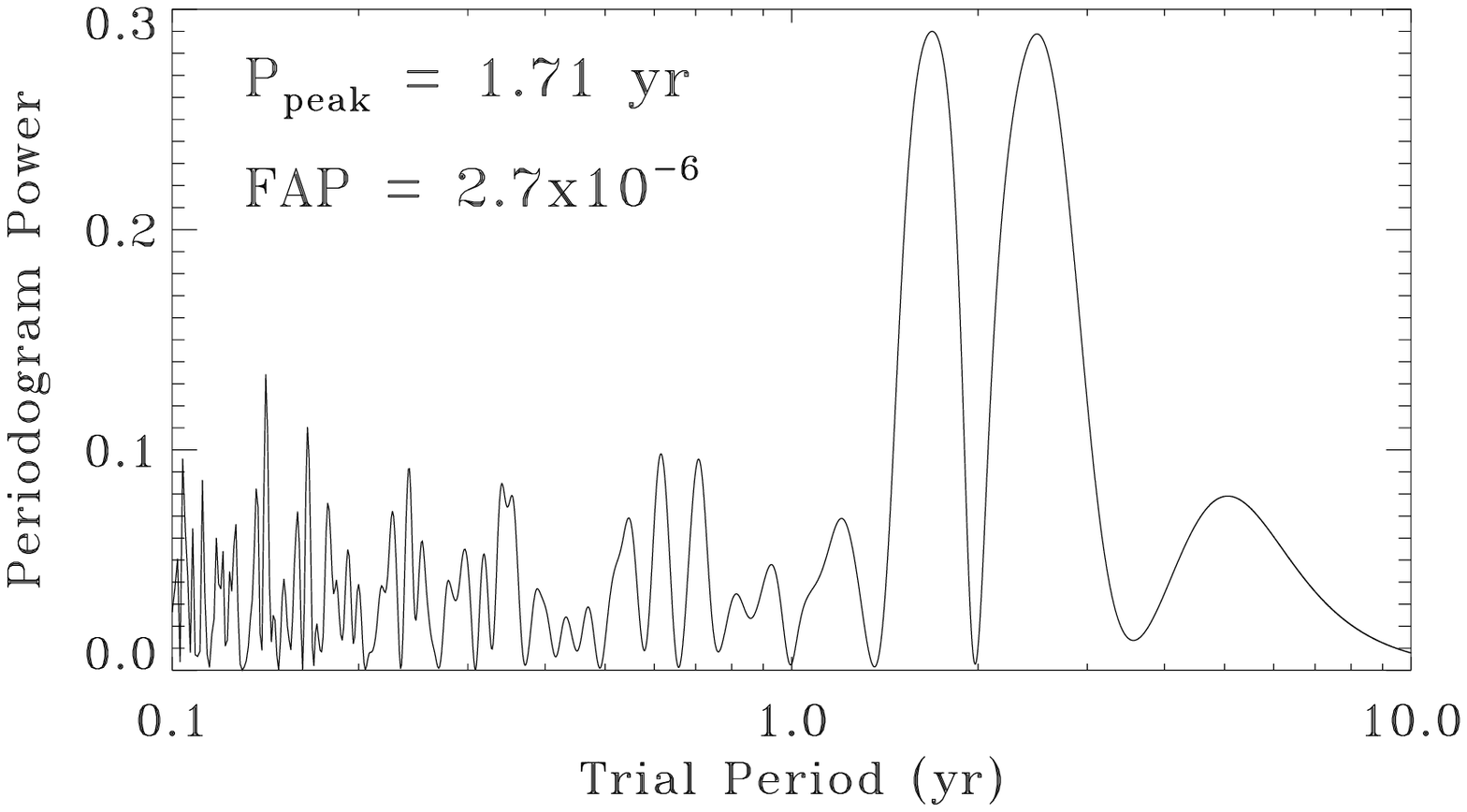} \\
\end{array}$    
\end{center}
\caption{Top: GLS periodogram of the $\xi$-axis residuals to the acceleration solution. Bottom: the same, for the $\eta$-axis.}    
\label{fig4}
\end{figure}

The addition of statistically significant acceleration terms, however, does not bring $\chi^2_r$ close to unity. One possibility for the excess residuals is that our 
formal per-measurement uncertainties in the astrometry, having a median of $\sim5$ mas, might be underestimated. We nevertheless investigated the residuals 
of the acceleration solution to see if any additional correlations could be identified, that might point to other sources of variability in our positional 
measurements. Fig. \ref{fig4} shows the results of the GLS periodogram analysis run on the post-acceleration fit residuals. While no significant power 
at any period is measured on the $\xi$-axis, a low-FAP (as determined via a standard bootstrap method) signal at $1.71\pm0.07$\,yr is clearly evident in the 
$\eta$-axis residuals (with a second peak of similar power at approximately 2.5\,yr). Quite interestingly, this value is very close 
to the $1.77\pm0.02$\,yr period clearly seen in the TOPP photometry, 
that corresponds to the well-known pulsation period of CW Leo (in the range $1.72-1.78$\,yr. See, e.g., \citealt{Lebertre1992}; \citealt{Kim2015}). 
Indeed, the $\eta$-axis residuals to the single-star (resp. acceleration) fit and the photometric time 
series show a rank-correlation coefficient of 0.53 (resp. 0.49), while no correlation is apparent between the photometry and the $\xi$-axis residuals. 

The excellent match between the periodicity seen in the astrometric residuals and the photometry of CW Leo calls for interpretation in terms of variability induced motion (VIM).  
This is an observable effect in astrometric measurements caused by brightness variations in one of the components of an unresolved double source or blended image, 
which manifest themselves as a strongly correlated shift of the optical photocenter. VIM effects, suggesting the presence of a binary companion, have been detected in Hipparcos 
observations of variable stars, especially long-period Miras (e.g. \citet{Pourbaix2003}, and references therein) and, more recently, in extensive re-analyses of 
Kepler photometry and astrometry \citep{Makarov2016}. 


We used the VIM modeling approach described in \citet{Wielen1996} to gain further insight on the possible architecture of the putative binary companion 
to CW Leo. As the VIM shows clearly only as a higher-order effect, we first modeled the residuals to the acceleration solution assuming a binary with a fixed relative geometry, 
described by a linear model with two parameters. 
This allows to derive the position angle $\Theta$ and a lower limit to the angular distance $\varrho$ between the components (for details, see \citealt{Wielen1996}). 
We obtain $\Theta$=$-12.7\pm2.1$\,deg (with $\Theta$ measured from North through East), and $\varrho\gtrsim 44$\,mas. 

\begin{figure}
 \begin{center}
$\begin{array}{c}
    \includegraphics[width=0.95\columnwidth]{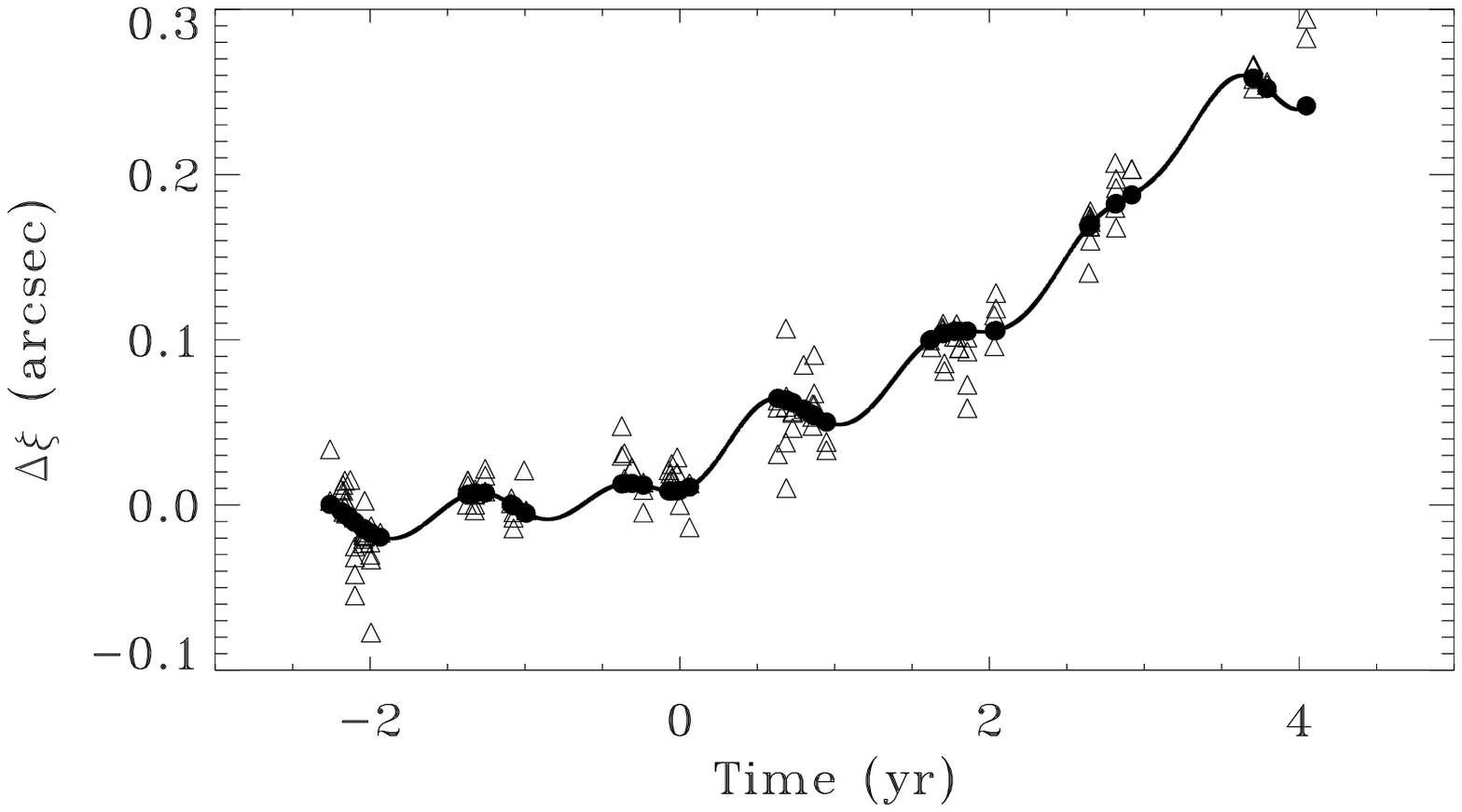} \\
    \includegraphics[width=0.95\columnwidth]{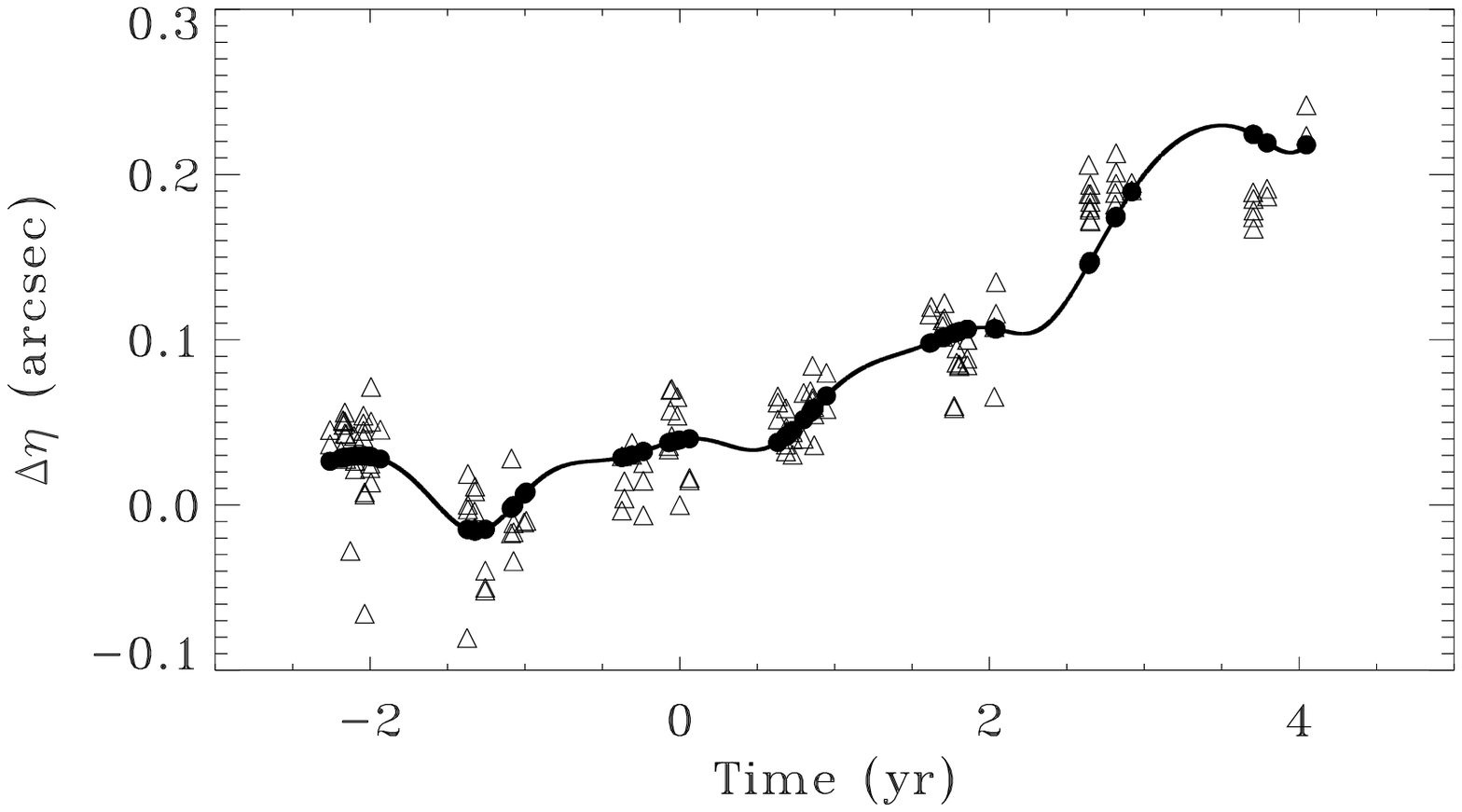} \\
\end{array}$    
\end{center}
\caption{Top: $\xi$ coordinate vs. time. The best-fit VIM+acceleration solution is superposed. Time is relative to the mean epoch (see Table \ref{sol}). 
Bottom: the same, for the $\eta$ coordinate.}
\label{fig6}
\end{figure}

We then used the \citet{Wielen1996} formalism in the case of 
binaries with accelerated motion. In this case, we model simultaneously astrometry and photometry to derive position offsets (in rectangular coordinates), 
proper motion ($\mu_{\alpha^\star,B}$ and $\mu_{\delta,B}$) and parallax $\pi_\mathrm{abs,B}$ of the center of mass of the binary (five linear parameters), 
position offsets (in rectangular coordinates as measured from the barycentre), 
orbital proper motion ($\mu_{\alpha^\star,1}$ and $\mu_{\delta,1}$) in rectangular coordinates moving with the barycentre), and the acceleration ($\dot\mu_{\alpha^\star,1}$ and 
$\dot \mu_{\delta,1}$) in the coordinate sytem of the barycentre of the variable component 
(six linear parameters), plus the total magnitude $m_\mathrm{c.o.m.}$ of the system (one non-linear parameter) when the photocentre coincides with the center-of-mass of the pair. 
We searched for the minimum of $\chi^2_r$ for the VIM model with accelerated motion as a function of a dense grid of input values for $m_\mathrm{c.o.m.}$. 
Table \ref{solvim} reports the best-fit solution ($\chi^2_r\simeq15.5$) obtained using a local minimization 
procedure (Levemberg-Marquardt). Based on an $F$-test, the VIM model with accelerated motion appears to be superior to the 
acceleration-only model (probability that the acceleration model better describes the data of $3.0\times10^{-9}$). The best-fit VIM+acceleration model 
is shown superposed to the data in Fig. \ref{fig6}. Attempts at fitting a full orbital VIM model based on a dense 
grid of trial periods exceeding the timespan of the observations did not produce an improvement in $\chi^2_r$. The rms of the residuals to the VIM+accelerated motion model 
($\sim20$ mas) still significantly exceeds the typical formal errors in the astrometry. However, a GLS analysis of the residuals to the VIM+acceleration fit did not indicate any 
significant periodicity, and no correlation was found between the residuals and the photometry. The secondary peak at $\sim2.5$\,yr initially seen in the $\eta$-axis 
residuals of the acceleration fit is thus to be interpreted as an alias of the main pulsation period. A typical long-term astrometric accuracy at the $15-20$ mas level 
might be more representative of the quality of our astrometric measurements, and would bring $\chi^2_r$ close to unity. We also note that the improved fit to the data reduces 
the parallax error by more than a factor of 2 with respect to the single-star model, bringing it in line with the typical precision of parallax measurements ($2-3$ mas) 
as determined for other TOPP targets in the past (e.g., \citealt{Smart2003}).

\begin{table}
\centering
\caption{VIM solution with accelerated motion for CW Leo. Uncertainties are derived via bootstrap method.}
\begin{tabular}{lr}
\hline  \\
 offset in $\alpha_B$ [mas]         & $41.13\pm2.59$  \\ 
 offset in $\delta_B$ [mas]                            &  $16.45\pm2.61$ \\ 
 $\mu_{\alpha^\star,B}$  [mas yr$^{-1}$]             &  33.38 $\pm$  1.60 \\ 
 $\mu_{\delta,B}$  [mas yr$^{-1}$]            &  25.43 $\pm$  1.69  \\ 
 $\pi_\mathrm{abs,B}$ [mas]                &  10.56 $\pm$  2.02\\    
 offset in $\alpha_1$ [mas]         & $-33.39\pm2.41$  \\ 
 offset in $\delta_1$ [mas]                            &  $27.55\pm2.42$ \\ 
 $\mu_{\alpha^\star,1}$  [mas yr$^{-1}$]             & $-$2.98 $\pm$  1.66 \\ 
 $\mu_{\delta,1}$  [mas yr$^{-1}$]            & $-$2.49 $\pm$  1.75  \\ 
 $\dot \mu_{\alpha^\star,1}$  [mas yr$^{-2}$]             &  24.24 $\pm$  1.91\\ 
 $\dot \mu_{\delta,1}$  [mas yr$^{-2}$]            &  21.69 $\pm$  1.89 \\ 
 $m_{c.o.m.}$  [mag]            &  11.24 $\pm$  0.16 \\ 

\hline 
\end{tabular}
\label{solvim}
\end{table}

\section{Discussion}

The analysis of $>6$\,yr of unpublished ground-based astrometric measurements from the TOPP program allows us to measure the trigonometric 
parallax of CW Leo (IRC+10216), the closest carbon-rich star to the Sun nearing the end of its AGB lifetime and characterized by extreme mass-loss. 
The direct distance estimate sits at the lower end of the range of indirect determinations reported in the recent literature (e.g. \citealt{Groenewegen2012}). 
Most importantly, we find convincing evidence for the presence of a binary companion to CW Leo, based on the detection of significant curvature in the 
residuals to a single-star model and on the identification of VIM effects that are successfully modelled simultaneously in astrometry and photometry. Any inferences that 
can be drawn on the orbital architecture (position angle, separation, orbital period) and mass of the companion ought to be seen in the 
context of the growing evidence for the existence of such an object. 

To our knowledge, the first claim of binarity for CW Leo dates back to the work of \citet{Guelin1993}, who proposed the possible existence 
of a 1\,M$_\odot$ companion with a period of $\sim800$\,yr based on arcsecond-level displacement of a molecular shell from the expected position 
of CW Leo at mm wavelengths. The tentative identification of a low-mass stellar companion with inferred period in the range $200-800$\,yr, similar to that proposed by 
\citet{Guelin1993}, was recently announced by \citep{Kim2015} using HST archival data. However, the detection might be spurious \citep{Stewart2016}. 
The innermost regions (a few arcsec) of the enviroment surrounding CW Leo have recently been probed with ALMA. Hints of rotating spiral structures 
in the inner wind envelope have been detected by D15, \citet{Cernicharo2015}, and \citet{Quintana2016}. Such structures can 
be interpreted in terms of the presence of a binary companion, but the exact geometry of the spirals is still uncertain. This has important 
repercussions on the possibility to constrain orbital elements (position angle, separation, eccentricity) and mass of the putative companion. 
For example, edge-on spirals structures induced by a K- or M-type companion star might place the secondary at $\sim20-25$ AU from CW Leo (depending on the 
adopted mass for the primary), with an approximate period of 55\,yr (D15), but if the spiral structure is seen closer to face-on 
the inferred separation might be much larger ($\sim65$ AU), for a similar companion mass (\citealt{Cernicharo2015}; \citealt{Quintana2016}). 

The results presented here lend further support to the existence of a stellar companion to CW Leo and help in constraining the range of possible orbital architectures 
and mass ratio of the system. The VIM analysis allows us to infer a small negative value (assuming fixed geometry) for the position angle ($\Theta$ = $-12$ deg) 
some 15 years before the ALMA observations presented 
by D15, who argue for $\Theta \sim 20$ deg. The direction of the accelerated motion of the primary agrees with the companion having moved eastward. 
The combined effect of the presence of the companion might also help to explain why the $\mu_{\alpha^\star}$ value derived in this work agrees with 
recent estimates obtained at mm wavelengths \citep{Menten2012}, while a statistically significant difference in $\mu_{\delta}$ with the \citet{Menten2012} value is observed. 
The discrepancy might be due to confusion between the VIM (contributing mostly in declination) and the proper motion. It then 
remains to be seen what inferences can be made on the mass and period of the companion based on the magnitudes of the perturbation and VIM effect. 

On the one hand, the value of $m_\mathrm{c.o.m.}$ we derive is proportional to that of the (assumed) constant secondary via the quantity ($1$+$q$)/$q$, 
with $q$ = $M_2/M_1$ being the binary mass ratio. 
An amplitude of the VIM effect of tens of mas is most likely obtained if the magnitude difference between the two components is not large 
($\Delta m\lesssim2$ mag, \citealt{Wielen1996}), thus, depending on the assumed mass for the primary, a late-type dwarf companion ($q\sim0.25-0.40$) at $\sim100$\,pc 
would roughly fit the scenario outlined by D15. 
On the other hand, at a distance of $\sim 80-110$\,pc as determined from our VIM+accelerated motion solution, the expected angular semi-major axis of the orbital motion of 
CW Leo around the barycenter would be $\alpha\sim70-125$ mas for a 1\,$M_\odot$ secondary at 20-25 AU, assuming a primary mass of 2.5\,$M_\odot$, and a circular orbit (D15). 
The variation of the position angle of the primary over the timespan of the TOPP observations (as derived from the VIM+acceleration model) is $\sim17$ deg, implying 
an orbital period $P\approx130$ yr, which is approximately compatible with the excursion of $\sim35$ deg over $\sim15$ yr between the $\Theta$ value obtained with the 
fixed-geometry VIM model and the one argued for by D15. 

The amount of curvature in the astrometric residuals to a single-star fit, with an excursion of some 50 mas over six years, would suggest that a significant 
fraction ($10-20\%$) of the orbital period has been covered by the TOPP observations. This can be made compatible with the tentative $P$ estimate 
above if the perturbation size were to be increased due to a lower primary mass. Indeed, \citet{Decin2011} argue for an envelope mass of CW Leo of $\sim2$\,M$_\odot$, 
thus the actual primary mass today might be around $1.5-2$\,M$_\odot$.  A value of $\alpha\approx120-200$ mas 
would then be more compatible with the inferred value of $P$, and would indicate an instantaneous orbital proper motion $2\pi\alpha/P\sim5-7$ mas yr$^{-1}$, 
in agreement within $1\,\sigma$ with that derived in our VIM+acceleration solution.
  
In summary, our results support the D15 evidence for the existence of a low-mass stellar companion to CW Leo with an orbital period shorter than those 
($\sim200-1000$\,yr) derived for other carbon-rich AGB stars, for which a binary companion is thought to be the cause of the detected spiral arm structure. 
However, our results would tend to indicate a longer orbital period (by a factor of two or so) than that tentatively reported by D15. The TOPP observations 
cannot directly constrain the orbital eccentricity, that could very well be significant as a consequence of the mass loss and mass exchange between the binary components 
(e.g., \citealt{Bonacic2008,Kim2017}).

Future astrometric and photometric measurements with Gaia will help tremendously in furthering our understanding of this remarkable system, but the interpretation of 
Gaia's space-borne astrometry of CW Leo will be improved significantly by the combination with the observations presented here, taking advantage of a time baseline of 
over 25 years. 

\section*{Acknowledgements}

P.G. gratefully acknowledges support from INAF through the "Progetti Premiali" funding scheme of the Italian Ministry of Education, University, and Research. 
We thank U. Abbas, J.-L. Halbwachs, P. Lucas, and E. Poretti for useful discussion. An anonymous referee provided insightful comments that helped to improve the manuscript.

\bsp	
\label{lastpage}
\end{document}